\documentclass[a4paper,11pt]{article}
\pdfoutput=1
\usepackage{jheppub}

\usepackage{epsf}
\usepackage{epsfig}
\usepackage{subfigure}
\usepackage{mathtools}
\usepackage{hhline}
\usepackage{float}
\usepackage{multirow}
\usepackage{nicefrac}
\usepackage{epstopdf}
\usepackage{slashed}
\usepackage{xcolor}
\usepackage{url}
\usepackage{multicol}
\usepackage{array}
\usepackage[normalem]{ulem}
\usepackage{graphicx}
\usepackage{umoline} 
\usepackage{color}
\allowdisplaybreaks

\newcommand{\be}{\begin{eqnarray}}
\newcommand{\ee}{\end{eqnarray}}

\newcommand{\ba}{\begin{array}}
\newcommand{\ea}{\end{array}}
\newcommand{\bee}{\begin{equation}\ba{c}}
\newcommand{\eee}{\ea\end{equation}}

\newcommand{\bi}{\begin{itemize}}
\newcommand{\ei}{\end{itemize}}

\title{Light quark loops in $K^\pm \to \pi^\pm \nu\bar\nu$ from vector meson dominance and update on the Kaon Unitarity Triangle}

\author{Enrico Lunghi$^1$,}
\author{Amarjit Soni$^2$}
\affiliation{
$^1$Physics Department, Indiana University, Bloomington, IN 47405, USA \\
$^2$Physics Department, Brookhaven National Laboratory, Upton, NY 11973, US\\
}
\emailAdd{elunghi@iu.edu} 
\emailAdd{adlersoni@gmail.com}

\abstract{
We use vector meson dominance to calculate non-perturbative contributions to the branching ratio of the rare decay $K^\pm \to \pi^\pm \nu\bar \nu$ stemming from matrix elements involving up-quark loops. The importance of this observable as well as of $K^0 \to \pi^0 l^+ l^-$ and of the direct CP violation parameter $\epsilon_K^{\prime}$ is then discussed in the context of a Unitarity Triangle sqtudy based on Kaon sector observables only.
}

\begin{document}

\maketitle 

\section{Introduction}
In recent years there has been considerable progress in understanding several long-standing tensions between experiments and Standard Model predictions. The anomalous magnetic moment of the muon has been measured by the Muon $g-2$ Collaboration~\cite{Muong-2:2024hpx} with an uncertainty which is about a third of the original E821 measurement~\cite{Muong-2:2006rrc}; at the same time recent progress on lattice QCD calculations of the hadronic vacuum polarization at low-$q^2$~\cite{RBC:2018dos, Borsanyi:2020mff, Ce:2022kxy, ExtendedTwistedMass:2022jpw, FermilabLattice:2022izv} are in tension with the corresponding results extracted from $e^+e^-$ data~\cite{Aoyama:2020ynm, Colangelo:2022jxc} and point to a reduced discrepancy between Standard Model (SM) and experimental determinations of the muon $g-2$. In the bottom sector, some anomalies in exclusive $b\to s\mu^+\mu^-$ decays persist albeit without tensions in the $\mu/e$ lepton universality ratios~\cite{Hurth:2021nsi, Greljo:2022jac, Ciuchini:2022wbq, Alguero:2023jeh} (see also ref.~\cite{Capdevila:2023yhq} for a review of and comparison between the results of various groups). Additionally, there are long-standing anomalies in $b\to c\ell\nu$ decays (see ref.~\cite{Reiss:2024zwb} for a short review and future experimental prospects). Standard Unitarity Triangle fits~\cite{UTfit:2022hsi, Qian:2023eok} do not show any significant tension with the CKM matrix~\cite{Cabibbo:1963yz, Kobayashi:1973fv} paradigm of flavor violation; note that the ongoing tension between inclusive and exclusive determinations of $|V_{cb}^{}|$ and $|V_{ub}^{}|$, while worrisome, is almost certainly not caused by new physics and will most likely find 
resolution in improved understanding of non-perturbative inputs (e.g. $B\to (\pi, D^{(*)})$ form factors, $B$-meson shape function, matrix elements of higher dimensional HQET operators, etc.). More recently the Belle-II collaboration presented a first evidence for the rare decay $B^+ \to K^+ \nu\bar \nu$~\cite{Belle-II:2023esi} which is about three sigma larger than the current SM prediction (see, for instance, ref.~\cite{Parrott:2022zte}). 

On this backdrop there is a set of quantities in the Kaon sector for which theoretical uncertainties are extremely small (or are expected to be further reduced in the near future) and that are the focus of an extensive experimental effort. See, for instance, ref.~\cite{Buras:2024mhy} for a clear and concise overview of the experimental situation, theoretical predictions within the SM and possible beyond-the-SM scenarios. Here we focus on the rare kaon decays $K^\pm \to \pi^\pm \nu\bar \nu$ and $K_L \to \pi^0\nu\bar \nu$, and on the direct CP violating quantity $\varepsilon^\prime/\varepsilon$. 

In particular, the charged mode branching ratio has been measured by the NA62 experiment~\cite{NA62:2021zjw} and found to be in agreement with state-of-art SM calculations~\cite{Brod:2021hsj}:
\begin{align}
{\rm BR} (K^+\to \pi^+\nu\bar \nu)_{\rm exp}\;  = &  
\left( 10.6^{+4.0}_{-3.4} |_{\rm stat} \pm 0.9_{\rm syst} \right) \times 10^{-11} \; , \\
{\rm BR} (K^+\to \pi^+\nu\bar \nu)_{\rm SM}\;  = &  
\left(7.73 \pm  0.16_{\rm pert} \pm 0.25_{\rm non-pert} \pm 0.54_{\rm param} \right) \times 10^{-11} \; ,
\end{align}
where the theoretical error breakdown corresponds to perturbative, non-perturbative, and parametric uncertainties. 
While the latter (dominated by the $|V_{ts}^* V_{td}^{}|$ combination of CKM matrix elements) are expected to be reduced in the future, the fate of the residual non-perturbative uncertainties due to up-quark loops is far from clear and is the focus of the present analysis. 

The current approach to long-distance contributions stemming from up-quark loops is based on the pioneer analysis presented in ref.~\cite{Isidori:2005xm}, where the authors integrate out the charm mass in perturbation theory and match the resulting $\Delta S = 1$ effective Lagrangian onto Chiral Perturbation Theory (ChiPT). The calculation of the contributions to the $K^+\to \pi^+\nu\bar\nu$ amplitude is then carried out at one-loop in ChiPT. Unfortunately, the calculation is incomplete because of lack of information on various $\Delta S = 1$ ChiPT counterterms. The latter are expected to be quite important numerically. For instance, in section~2.2 of ref.~\cite{DAmbrosio:1998gur} the authors discussed the $K^+\to \pi^+ e^+ e^-$ mode (for which the single photon exchange receives long-distance contributions that are very similar to the $\nu\bar\nu$ mode), showed that contributions from higher-mass intermediate states can be parameterized by a polynomial in $q^2$ (which receives contributions from the unknown ChiPT low-energy constants) and that the latter is numerically dominant.

In this work, we follow a different strategy and use resonant ChiPT~\cite{Ecker:1988te, Ecker:1989yg, Ecker:1992de}\footnote{See ref.~\cite{Park:2024mrw} for a different approach to the treatment of vector meson resonances in ChiPT.} to describe $Z$-boson mediated non-perturbative effects associated to up-quark loops. The basic idea is that pion loops can be described in terms of off-shell $\rho$ meson exchanges. In this approach the missing counterterms appear in the matching of the QCD $\Delta S = 1$ Lagrangian onto the Resonant ChiPT Lagrangian involving the $\pi$, $K$ and $\rho$ mesons. These Wilson coefficients can in turn be estimated using various approaches like the factorization~\cite{Pich:1991fq, Cheng:1990wx} or weak deformation~\cite{Ecker:1990in} models, albeit with large uncertainties. A similar approach has been applied to the calculation of non-perturbative effects in $B\to K \ell^+\ell^-$ in ref.~\cite{Isidori:2024lng}, where the focus is on charm loops and their description in terms of $D_{(s)}^{(*)}$ exchanges.

Finally we stress that a proper precise calculation of these non-perturbative effects is only possible in the lattice-QCD framework. Recent exploratory studies~\cite{Christ:2016lro, Bai:2017fkh, Bai:2018hqu, Christ:2019dxu, Blum:2022wsz} have shown that this calculation is possible but seems computationally intensive especially for the needed accuracy and  can hopefully be completed in the near feature.

The neutral $K_L \to \pi^0 \nu\bar\nu$ gold-plated mode is currently being studied by the KOTO experiment which published results based on data collected in 2015~\cite{KOTO:2018dsc}, 2016-2018~\cite{KOTO:2020prk} and 2019-2021~\cite{KOTOMoriond2024}\footnote{The analysis of data collected in 2019-2021 is still at a preliminary level.}. The current experimental upper limit and the corresponding SM prediction~\cite{Brod:2021hsj} are:
\begin{align}
{\rm BR} (K_L\to \pi^0\nu\bar\nu)_{\rm exp} \; & < 2.0 \times 10^{-9} \;\; @ 90\%\; {\rm C.L.} \; , \\
{\rm BR} (K_L\to \pi^0\nu\bar\nu)_{\rm SM} \; & = \left(2.59 \pm 0.29 \right) \times 10^{-11}  \; .
\end{align} 
Thus we see that the current KOTO experimental upper bound is about two orders of magnitude above the SM prediction. Moreover, to actually measure the SM parameters to some reasonable precision, requires observation of at least a handful of events. Looking back at the history of the somewhat experimentally  easier charged kaon mode $K^+ \to \pi^+ \bar \nu \nu$ (see figure~4 of ref.~\cite{Ceccucci:2021gpl}), one can easily see that such a progress could take a decade or even more. It is, therefore, important to constrain indirectly this gold plated mode as much as possible. A good way to do this may be via $K^0 \to \pi^0 l^+ l^-$\cite{Schacht:2023vsz}.  Experimentally this is a much easier mode. Besides, studies of $K_S$ at LHCb and $K_L$ at JPARC can provide valuable experimental information on CP conserving and CP violating decays. The theoretical challenge is to quantify precisely the relative size of the one photon versus the two photon mediated contributions to the $l^+ l^-$ mode, a task that can be addressed using ChiPT~\cite{Ecker:1987qi, DAmbrosio:1998gur, Isidori:2004rb, DAmbrosio:2018ytt, Buchalla:2003sj}, phenomenological modelling\cite{Donoghue:1987awa, Schacht:2023vsz} as well as lattice QCD calculations~\cite{Christ:2015aha, RBC:2022ddw}. 

Finally, we stress that new physics models that contribute to $K^\pm \to \pi^\pm \nu\bar\nu$ are also expected to contribute not only to the CP violating neutral mode $K_L \to \pi^0 \nu\bar\nu$, but also to $\varepsilon_K$ and $\varepsilon^\prime_K/\varepsilon_K$ which quantify indirect and direct CP violation in $K\to \pi\pi$ decays. In order to isolate new physics contributions to the Kaon sector it is useful to consider Unitarity Triangle analyses in which $B$ and $K$ observables are considered separately; in ref.~\cite{Lehner:2015jga}, we showed how the Kaon Unitarity Triangle (KUT) can be used for this purpose. Given the experimental and theoretical progress achieved in the last decade, we will conclude this study with an updated study of the present status of the KUT and of future prospects.

The paper is structured as follows. In section~\ref{sec:2} we introduce the effective Hamiltonian that controls $K^+\to \pi^+ \nu\bar\nu$ and discuss the perturbative part of the branching ratio calculation. In section~\ref{sec:currentstatus} we present the current status of the calculation of long distance contributions due to light-quarks loops. In section~\ref{sec:rolevector} we present a heuristic discussion of the role of vector meson exchange in the calculation of these non-perturbative effects. Sections~\ref{sec:vector} and \ref{sec:calculation} are the main part of this paper and summarize the calculation of these effects using resonant Chiral Perturbation Theory. In section~\ref{sec:kut} we present an updated study of the Kaon Unitarity Triangle. In section~\ref{sec:conclusions} we present our conclusions.

\section{\texorpdfstring{$K \to \pi \nu \bar \nu$}{K2pinunu} decays}
\label{sec:2}
The effective Hamiltonian which describes the decays $K^+ \to \pi^+ \nu\bar\nu$ and $K_L\to \pi^0 \nu\bar\nu$ in the Standard Model is:
\begin{align}
{\cal H}_{\rm eff} &\ni \frac{4 G_F}{\sqrt{2}} \left[
 \lambda_t \; C_\nu \; O_\nu + \sum_{\substack{i=1,2\\ q=u,c}}  \lambda_q \; C_i^q \; Q_i^q 
\right] \;.
\end{align}
The operators are defined as
\begin{align}
Q_{\nu} &= (\bar s_L \gamma_\mu d_L) (\bar \nu_L \gamma^\mu \nu_L)  \; , \\
Q_1^c &= (\bar s_L T^a \gamma^\mu u_L) (\bar u_L T^a \gamma_\mu d_L) \; , \\
Q_2^c &= (\bar s_L \gamma^\mu u_L) (\bar u_L \gamma_\mu d_L) \; , \\
Q_1^u &= (\bar s_L T^a \gamma^\mu u_L) (\bar u_L  T^a \gamma_\mu d_L) \; , \\
Q_2^u &= (\bar s_L \gamma^\mu u_L) (\bar u_L \gamma_\mu d_L) \; ,
\end{align}
where $\lambda_q = V_{qs}^* V_{qd}^{}$ are various combinations of CKM elements~\cite{Cabibbo:1963yz, Kobayashi:1973fv} and $T^a$ are the Gell-Mann matrices.

The SM expressions for the $K^+ \to \pi^+ \nu\bar\nu$ and $K_L\to \pi^0 \nu\bar\nu$  branching ratio are (see, for instance, ref.~\cite{Haisch:2007pd}):
\begin{align}
{\rm BR} (K_L \to \pi^0 \nu\bar\nu) = \; & 
\kappa_L \left( \frac{{\rm Im} \lambda_t}{\lambda^5} X(x_t) \right)^2 \; , \label{kl2pnn} \\
{\rm BR} (K^+ \to \pi^+ \nu\bar\nu) = \; & 
\kappa_+ (1+\Delta_{\rm EM}) \Bigg[ 
 \frac{{\rm Im} \lambda_t}{\lambda^5} X(x_t) \nonumber \\
& +
 \left(
 \frac{{\rm Re} \lambda_t}{\lambda^5} X(x_t) 
 +  \frac{{\rm Re} \lambda_c}{\lambda} (P_c + \delta P_{c,u})
  \right)^2\Bigg] \; ,
\label{kp2pnn}
\end{align}
%
where $\kappa_{L,+}$ are normalization factors, $\lambda$ is the Cabibbo angle, $X(x_t)$ is a loop-function known at NLO in QCD, $\Delta_{\rm EM}$ encapsulates QED corrections, $P_c$ originates from perturbative diagrams with internal charm quarks. References and numerical values for the above mentioned quantities are collected in table~\ref{tab:inputs}.

In eq.~(\ref{kp2pnn}), the charm quark has been integrated out completely. In a first step, contributions from scales above $m_c$ are calculated in terms of $Z$-penguin and $W$-box diagrams involving charm quarks. These contributions are known at NNLO, yield the quantity $P_c$ and correspond to charm effects on the Wilson coefficient of the operator $Q_{\nu}$ (see references listed in table~\ref{tab:inputs}). 

In addition, there are two subdominant effects corresponding to the contribution of dimension-eight operators generated $\mu = O(m_c)$ and of up-quark loops. The former are suppressed by $O(\Lambda_{\rm QCD}^2/m_c^2)$ compared to leading dimension-six operator. The latter are controlled by long distance effects and can be calculated by matching the Weak Effective Hamiltonian onto Chiral Perturbation (ChiPT). In ref.~\cite{Isidori:2005xm}, the authors present a detailed analysis of both of these terms and calculate up-quark loops contributions at tree and one-loop level in ChiPT (up to certain unknown counterterms). Effects from dimension-eight operators and up-quark loops are included in the quantity $\delta P_{c,u}$. In ref.~\cite{Isidori:2005xm} this quantity is calculated including only tree-level ChiPT contributions (the one-loop calculation is incomplete due to the missing counterterms) with an uncertainty estimated at the 50\% level: $\delta P_{c,u} = 0.04 \pm 0.02$. 

The focus of the first part of this work is to present an alternative calculation of the up-quark loops contribution. In the next sections we review the leading power matching of the Weak effective Hamiltonian onto ChiPT and explain how pseudoscalar loop effects can be replaced by a calculation involving vector meson exchanges in the framework of Resonant Chiral Perturbation Theory~\cite{Ecker:1988te, Ecker:1989yg}.

\begin{table}[t]
\begin{center}
\begin{tabular}{|l|l|l|}\hline 
$\lambda$ & $0.22519 \pm 0.00083 $ & \cite{UTfit:2022hsi} \cr
$\kappa_L$ & $(2.231 \pm 0.013)\times 10^{-11} (\lambda/0.2252)^8$ & \cite{Mescia:2007kn} \cr 
$\kappa_+$ & $(5.173 \pm 0.025)\times 10^{-11} (\lambda/0.2252)^8$ & \cite{Mescia:2007kn} \cr 
$X(x_t)$ & $1.462 \pm 0.017_{\rm QCD} \pm 0.002_{\rm EW}$ & 
\cite{Buchalla:1993bv, Buchalla:1998ba, Misiak:1999yg, Brod:2010hi, Buras:2015qea, Brod:2021hsj} \cr
$\Delta_{\rm EM}$ & $ -0.003$ & \cite{Mescia:2007kn} \cr
$P_c$ & $(0.3604\pm 0.0087)(0.2255/\lambda)^4$ & \cite{Buras:2006gb, Buchalla:1998ba, Buchalla:1993wq, Buras:2005gr, Isidori:2005xm, Brod:2008ss, Brod:2021hsj} \cr
$g_8$ & $3.58 \pm 0.14$ & \cite{Cirigliano:2019cpi} \cr 
\hline
\end{tabular}
\end{center}
\caption{Inputs used in the numerical analysis.   }
\label{tab:inputs}
\end{table}

\section{Current status of long distance contributions to \texorpdfstring{$K^+ \to \pi^+ \nu\bar\nu$}{K+2pi+nunubar}}
\label{sec:currentstatus}
In this section we present a review of non-perturbative up quark contributions to $K^+ \to \pi^+ \nu \bar \nu$ following closely the analysis presented in ref.~\cite{Isidori:2005xm} (see refs.~\cite{Rein:1989tr, Hagelin:1989wt, Lu:1994ww, Geng:1994cw, Geng:1995np, Fajfer:1996tc, Bijnens:1997vq, Bijnens:1998fm, Geng:1999pu} for previous studies along similar lines).

\begin{figure}[t]
\begin{center}
\includegraphics[height=0.2 \linewidth]{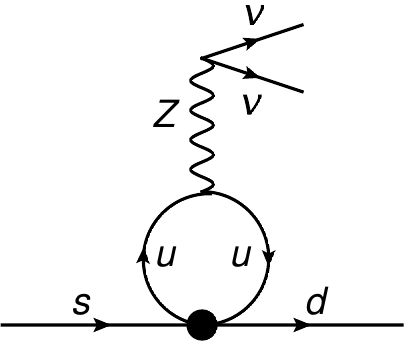} \\
\caption{Long distance contributions to $K^+\to \pi^+\nu\bar\nu$ at the quark level.}
\label{fig:longdistance-quark}
\end{center}
\end{figure}

The starting point is the Weak Effective Hamiltonian at $\mu \sim O(1\; {\rm GeV})$
\begin{align}
{\cal L}_{\rm eff}(\mu) &= 
{\cal L}_{\rm QCD} + \frac{4 G_F}{\sqrt{2}} \bar q \gamma^\mu (v_\mu + \gamma_5 a_\mu) q
- \frac{4 G_F}{\sqrt{2}} \lambda_u \sum_{i=1,2} C_i^u Q_i^u \; .
\label{eq:Leffmu1}
\end{align}
Within QCD, the contribution of the operators $Q_i^u$ starts at one-loop (see figure \ref{fig:longdistance-quark}) and is controlled by long-distance effects. To overcome this problem, the effective Lagrangian in eq.~(\ref{eq:Leffmu1}) is matched onto the $SU(3)$ Chiral Lagrangian in which the only degrees of freedom are the eight pseudoscalar mesons. The main ingredients are the quarks $SU(3)$ triplet is $q= (u,d,s)$ and the external vector and axial currents $v_\mu$ and $a_\mu$, which are given in terms of lepton fields and appear upon integrating out the $W^\pm$ and $Z$ vector bosons. 

For very small pseudoscalar meson momenta, one can follow the spurion approach described in ref.~\cite{Gasser:1984gg, Bernard:1985wf} which is based on the observation that ${\cal L}_{\rm eff}(\mu)$ is invariant under local $SU(3)_L\times SU(3)_R$ transformations of the quark triplet $q_{L,R}$ as long as the external currents (which in our case involve neutrino bilinears) are assigned appropriate gauge transformations. This local gauge symmetry is then preserved in ChiPT as long as a covariant derivative involving the external currents is used. 

At the $\Delta S=0$ level (first two terms in eq.~(\ref{eq:Leffmu1})) and at lowest order in the Chiral expansion, the Chiral Lagrangian contains a single operator:
\begin{align}
{\cal L}_{\rm ps}^{} &= 
\frac{F^2}{4} \langle D_\mu U D^\mu U^\dagger \rangle  \; , 
\label{Lps} 
\end{align}
where $F = 92.1(6)$ MeV~\cite{Aoki:2019cca} is the pion decay constant in the Chiral limit, $\langle\rangle$ is the trace over flavor $SU(3)$ indices and $U$ is the standard parameterization of the pseudoscalar mesons octet
\begin{align}
U &= u^2 = e^{i\sqrt{2} \Phi/F} \; , \label{eq:U}\\
\Phi &= 
 \begin{pmatrix}
\pi^0 /\sqrt{2} + \eta_{8}/\sqrt{6} & \pi^+ & K^{+} \cr
\pi^- & - \pi^0/\sqrt{2} + \eta_{8}/\sqrt{6} & K^{0} \cr
K^{-} & \bar K^{0} & -2 \eta_{8}/\sqrt{6}\cr
\end{pmatrix}  \; .
\end{align}
The external currents enter via the covariant derivative~\cite{Isidori:2005xm}:
\begin{align}
D_\mu &= \partial_\mu U - i g_Z Z_\mu \left(\sin^2\theta_W [Q,U] + UQ -\frac{a_1}{6} U \right) 
-i \frac{g}{\sqrt{2}} U \left( T_+ W_\mu^+ + {\rm h.c.} \right) \; , \\
Z_\mu &= \frac{g_Z}{4 M_Z^2} \sum_{\ell} \bar \nu_\ell \gamma_\mu (1-\gamma_5)\nu_\ell \; ,
\label{nc}\\
W_\mu^+ &= \frac{g}{2\sqrt{2} M_W^2} \sum_\ell \bar \ell \gamma_\mu (1-\gamma_5) \nu_\ell \, \label{cc} 
\end{align}
where $Q = \text{diag}(2/3,-1/3,-1/3)$ and $g_Z = g/\cos\theta_W$. The $a_1$ term corresponds to the matching of the singlet component of the left-handed weak current. The anomaly in the $U(1)_L$ current is responsible for the deviation of $a_1$ from 1. In the $N_c\to \infty$ limit, the effect of the anomaly vanishes and, correspondingly, $a_1\to 1$~\cite{Lu:1994ww}.  

The $\Delta S=1$ term in ${\cal L}_{\rm eff}(\mu)$ (last term in eq.~(\ref{eq:Leffmu1})) matches onto the only $\Delta S=1$ operator which appears at lowest order in ChiPT:
\begin{align}
{\cal L}_{\rm ps}^{\Delta S=1} &= G_8\; F^4 \; \langle \lambda_{sd} \left[
D^\mu U^\dagger D_\mu U -2 i g_Z Z_\mu U^\dagger D^\mu U \left(Q-\frac{a1}{6} \right)\right] \rangle,
\label{LpsDS1}
\end{align}
where $G_8 = -  V_{ud}^{} V_{us}^* G_F/\sqrt{2}  \; g_8$ with $g_8 = 3.58 \pm 0.14$~\cite{Cirigliano:2019cpi},
and $\lambda_{s d} = (\lambda_6-i \lambda_7)/2 = \delta_{i3}\delta_{j2}$ ($\delta_{ij}$ is the Kronecker delta) is a combinations of Gell-Mann matrices which singles out the $\bar s \to \bar d$ transition. At the quark level, these contributions correspond to diagrams in which the weak $\Delta S=0$ currents attach to the external legs of the $\Delta S=1$ operators $Q_{1,2}^u$.

In this framework, long distance contributions originating from up-quark loops are obtained, at order  $O(p^2)$, by calculating the $K^+\to \pi^+\nu\bar\nu$ amplitude using ${\cal L}_{\rm ps}^{\Delta S=0} + {\cal L}_{\rm ps}^{\Delta S=1}$. There are four tree--level diagrams that contribute (see figures~2 and 5 of ref.~\cite{Isidori:2005xm}). In three of them, the $\Delta S=1$ transition happens at the $Z$ current vertex (one insertion of ${\cal L}_{\rm ps}^{\Delta S=1}$) or on the external $K^+$ and $\pi^+$ lines (one insertion of ${\cal L}_{\rm ps}^{\Delta S=1}$ and of ${\cal L}_{\rm ps}^{\Delta S=0}$ each); in the fourth diagram two insertions of the charged current $W_\mu^+$ in eq.~(\ref{cc}) yield a $t$--channel contribution with an intermediate charged lepton. The resulting contribution to the phase space averaged $K^+\to \pi^+\nu\bar\nu$ amplitude is
\begin{align}
\delta P_{c,u} &= \frac{1}{3} \sum_{\ell=e,\mu} \langle P_Z(q^2) + P_{WW}^\ell (q^2) \rangle 
= \frac{\pi^2 F^2}{\lambda^4 M_W^2} \left[ \frac{4 |G_8|}{\sqrt{2} \lambda G_F} -\frac{4}{3} \right] \; ,
\label{ps-tree}
\end{align}
where $\lambda$ is the Cabibbo angle. 

In ref.~\cite{Isidori:2005xm}, the authors present a complete discussion of one-loop corrections to eq.~(\ref{ps-tree}) but refrain to use the latter in the numerics because of the lack of knowledge about certain $O(p^4)$ counterterms. eq.~(\ref{ps-tree}) is the expression currently used to include the effects of non-perturbative up-quark loops to $K^+\to \pi^+\nu\bar\nu$ and yields
\begin{align}
\delta P_{c,u} = 0.03\; (1 \pm 0.5)\; ,
\label{eq:GinoUpdated}
\end{align}
where the 50\% uncertainty is assigned in order to capture the potential size of the missing $O(p^4)$ corrections. The central value in eq.~(\ref{eq:GinoUpdated}) is slightly lower than the result quoted in ref.~\cite{Isidori:2005xm}, $ 0.04 \pm 0.02$, mainly because of the updated input value for the matching coefficient $G_8$: as discussed below eq.~(\ref{LpsDS1}), we adopt $|G_8| =  |V_{ud}^{} V_{us}^*| G_F/\sqrt{2} g_8 = (6.7 \pm 0.3 ) \times 10^{-6}\; \text{GeV}^{-2}$ while in ref.~\cite{Isidori:2005xm} the value $|G_8| \simeq  9 \times 10^{-6}\; \text{GeV}^{-2}$ was used.

\section{The role of vector mesons exchange}
\label{sec:rolevector}
One issue that, to the best of our knowledge, has not been addressed in the context of nonperturbative up-quark loop contributions to $K^+\to \pi^+\nu\bar\nu$ is the role of vector mesons.

It is well known that the chiral expansion breaks down at large momentum transfer. ChiPT becomes completely nonperturbative at scales of order $\sqrt{s} \sim 4\pi F \simeq O(1.2\; {\rm GeV})$ but large deviations from calculations at order $O(p^2,p^4)$ may be expected already at much smaller $\sqrt{s}$ values.

For instance, a clear example of the failure of the perturbative ChiPT expansion is offered by the pion form factor. In figure 1 of ref.~\cite{Gonzalez-Solis:2019iod} the authors compare the ChiPT calculation at order $p^4$ with the measured form factor. It is clear that the perturbative expansion at order $p^4$ deviates by 50\% at $\sqrt{s} \sim 400\; {\rm MeV}$. At larger $\sqrt{s}$ values, the form factor is dominated by $t$-channel exchange of a $\rho$ meson and displays a typical Breit-Wigner behavior (see refs.~\cite{Kroll:1967it, Lee:1967iv, Gounaris:1968mw, Cho:1969xza} for early studies of vector mesons role in the pion form factor).
%
%

Another enlightening example of vector meson dominance is offered by an analysis of contributions of the leading ChiPT Lagrangian to the several low-energy constants which appear in ChiPT at order $p^4$. While ChiPT is non-renormalizable and these constants have to be ultimately extracted from measurements, it is interesting to calculate the contributions from one-loop pseudoscalar exchange and compare them to tree-level vector meson contributions. This study has been presented in  ref.~\cite{Ecker:1992de} (where vector mesons interactions have been included following the approach of refs.~\cite{Ecker:1988te, Ecker:1989yg}). The main result is that vector meson contributions alone are sufficient to reproduce the observed values of all low-energy constants; suggesting that vector meson exchange diagrams dominate over renormalized one-loop contributions. 

A final example of vector meson dominance over pseudoscalar loops is offered by recent lattice QCD calculations of the hadronic contribution to the photon vacuum polarization in the context of the prediction for the muon anomalous magnetic moment (see, for instance, refs.~\cite{Chakraborty:2015ugp, Chakraborty:2016mwy, Colangelo:2018mtw, Aoyama:2020ynm}). 

The above remarks apply to the matching of the the $\Delta S = 1$ weak operator which couple directly to the $Z$-boson mediated neutrino current. The contributions from $W$-boson box diagrams do not correspond to the exchange of any spin one resonance; therefore, within chiral perturbation theory, beyond tree level they are described by the one-loop diagrams discussed in ref.~\cite{Isidori:2005xm}.

The point of view that we adopt is that, while it is reasonable to use the ChiPT Lagrangian in eq.~(\ref{LpsDS1}) to describe tree-level and one-loop effects corresponding to the $W$ box diagram, $u\bar u$ loops contributions to the $K^+\to \pi^+\nu\bar\nu$ amplitude which couple to the $Z$ mediated current are better described by tree-level vector meson exchanges.

On one hand, it is clear that, for $q^2 = (p_K-p_\pi)^2$  near the lowest lying vector meson ($\rho$), ChiPT breaks down completely; on the other one, the kinematics of the decay $K^+\to \pi^+ \nu\bar\nu$ implies $q^2/m_\rho^2 < (m_K-m_\pi)^2/m_\rho^2 \simeq 0.21$. The question whether, in this $q^2$ range, vector meson exchanges should completely replace pseudoscalar loops or be included alongside with, is not settled yet. 

Schematically the amplitude can be decomposed as
\begin{align}
A(q^2) = \left[A(0)\right]_{\text{tree-level}} + \left[A(q^2)-A(0)\right]_{\rho \text{-exchange}} +  \left[A(q^2)-A(0)\right]_{\text{non-resonant}}\; .
\end{align}
In this paper, we focus on the calculation of the $\rho$-exchange contribution.

In the next section we follow the approach of refs.~\cite{Ecker:1988te, Ecker:1989yg}, known as Resonant Chiral Perturbation Theory, to include vector mesons in the Chiral Lagrangian. An alternative description of vector mesons interactions is provided by so-called Hidden Local Symmetry approach (see ref.~\cite{Harada:2003jx} for a review) which has been shown to be equivalent to the Resonant ChiPT description. We chose the Resonant ChiPT formalism because it allows, in our opinion, for a more straightforward matching of $\Delta S=1$ operators. Obviously, the extraction of the relevant Wilson coefficients is dominated by non-perturbative physics and we will have to rely on some more-or-less naive estimates based on naive factorization or the heuristic weak deformation model~\cite{Ecker:1992de}.

The basic strategy follows the same steps as in section~\ref{sec:currentstatus}. We first match the first two terms of eq.~(\ref{eq:Leffmu1}) onto a Chiral Lagrangian that describes $\Delta S=0$ vector mesons couplings to pseudoscalar mesons and to the external currents $Z_\mu$ and $W^+_\mu$ in eqs.~(\ref{nc}) and (\ref{cc}). Then we match the final  term in eq.~(\ref{eq:Leffmu1}) onto a purely hadronic $\Delta S=1$ Chiral Lagrangian (which contains, for instance, the $\rho$-$K$-$\pi$ vertex). 

In this framework corrections to tree-level ChiPT effects are given by the diagrams in figure~\ref{fig:longdistance-chipt}. Solid (blue) circles are insertions of the $\Delta S=1$ pseudoscalar Lagrangian, empty black squares are insertions of $\Delta S=0$ vector-pseudoscalar mesons vertices, solid (blue) squares are $\Delta S=1$ vector-pseudoscalar mesons vertices, and small (black) circles are the couplings between vector mesons and the external currents.

\section{Chiral Lagrangian for pseudoscalar and vector mesons}
\label{sec:vector}

\subsection{\texorpdfstring{$\Delta S = 0$}{Delta S = 0}}
\label{sec:vectorDS0}
The pseudoscalar meson Chiral Lagrangian in presence of external current is given in eq.~(\ref{Lps}). We include vector mesons into the Chiral Lagrangian following the approach and notation of refs.~\cite{Ecker:1988te, Ecker:1989yg}. We choose to work with antisymmetric rank-two tensor fields ($V_{\mu\nu}$) rather than massive vector fields ($V_\mu$). It is clear that, at a given order in the Chiral expansion, the latter approach, as discussed in ref.~\cite{DAmbrosio:2006xmn}, involves more operators (in both cases the fields have mass dimension 1 but the antisymmetric fields have two Lorentz indices). Nevertheless, an explicit analysis presented in ref.~\cite{Ecker:1989yg}, demonstrated that the two formulations yield identical results as long as appropriate matching conditions on certain operators appearing in the $O(p^4)$ pseudo-scalar meson Lagrangian are imposed. These matching conditions correspond precisely to the counter-terms that are missing in the $O(p^4)$ calculation discussed in ref.~\cite{Isidori:2005xm} and that can, therefore, be estimated by vector mesons exchanges in the antisymmetric field formalism.

The antisymmetric fields corresponding to the vector mesons transform as an octet of isospin $SU(3)$:
\begin{align}
\hat V_{\mu\nu} &= 
\begin{pmatrix}
\rho^0_{\mu\nu} /\sqrt{2} + \omega_{8{\mu\nu}}/\sqrt{6} & \rho^+_{\mu\nu} & K^{*+}_{\mu\nu} \cr
\rho^-_{\mu\nu} & - \rho^0_{\mu\nu}/\sqrt{2} + \omega_{8{\mu\nu}}/\sqrt{6} & K^{*0}_{\mu\nu} \cr
K^{*-}_{\mu\nu} & \bar K^{*0}_{\mu\nu} & -2 \omega_{8{\mu\nu}}/\sqrt{6}\cr
\end{pmatrix} 
+ \frac{\omega_{1{\mu\nu}}}{\sqrt{3}} \; \mathbb{I}_{3\times 3}  \; .
\end{align}
where~\cite{Guo:2008sh}
\begin{align}
\omega_{1{\mu\nu}} = \sqrt{\frac{2}{3}} \omega_{\mu\nu} - \sqrt{\frac{1}{3}}  \phi_{\mu\nu} \; , \\
\omega_{8{\mu\nu}} = \sqrt{\frac{2}{3}} \phi_{\mu\nu} + \sqrt{\frac{1}{3}}  \omega_{\mu\nu} \; .
\end{align}

The leading power $\Delta S=0$ vector meson Lagrangian including interactions with external currents is given by (see refs.~\cite{Ecker:1988te, Ecker:1989yg})\footnote{The same Lagrangian expressed in terms of vector fields $V_\mu$ involve two additional operators, $\langle V_\mu [u_\nu,f_-^{\mu\nu}]\rangle$ and $\langle V_\mu [u^\mu,u^\dagger m u^\dagger - u m^\dagger u]\rangle$ as discussed in ref.~\cite{DAmbrosio:2006xmn}.}:
\begin{align}
{\cal L}_{\rm vect} &= 
-\frac{1}{2} \langle \nabla^\lambda \hat V_{\lambda\mu} \nabla_\nu \hat V^{\nu\lambda} -\frac{1}{2} \hat M_V^2\; \hat V_{\mu\nu} \hat V^{\mu\nu}\rangle  +\frac{1}{2\sqrt{2}} \left[
F_V \langle \hat V_{\mu\nu} f_+^{\mu\nu} \rangle 
+ i G_V \langle \hat V_{\mu\nu} [u^\mu, u^\nu] \rangle
\right] \; ,
\label{eq:vect}
\end{align}
where 
\begin{align}
\nabla_\lambda \hat V_{\mu\nu} &= \partial_\lambda \hat V_{\mu\nu} + \left[ \Gamma_\lambda, \hat V_{\mu\nu} \right] \;  , \\
\Gamma_\mu &= \frac{1}{2} \left[ u^\dagger (\partial_\mu - i r_\mu) u + u (\partial_\mu - i l_\mu) u^\dagger \right] \; , \\
u_\mu & =  i u^\dagger (D_\mu U) u^\dagger = u_\mu^\dagger\; , \\
f^{\mu\nu}_\pm & = u F_L^{\mu\nu} u^\dagger \pm u^\dagger F_R^{\mu\nu} u \; , \\
F_{R,L}^{\mu\nu} &= \partial^\mu (v^\nu \pm a^\nu) - \partial^\nu (v^\mu \pm a^\mu) - i [v^\mu \pm a^\mu, v^\nu \pm a^\nu] \; , \\
v_\mu &= -e Q A_\mu -\frac{g}{2\cos\theta_W} \left[ Q \cos(2 \theta_W) -\frac{1}{6} \right] Z_\mu
-\frac{g}{2\sqrt{2}} (T_+ W_\mu^+ + \text{h.c.} ) \; , \\
a_\mu &= \frac{g}{2\cos\theta_W} \left[ Q  -\frac{1}{6} \right] Z_\mu
+\frac{g}{2\sqrt{2}} (T_+ W_\mu^+ + \text{h.c.} ) \; , \\
T_+ &= \begin{pmatrix} 0 & V_{ud} & V_{us} \cr 0 & 0 & 0 \cr 0 & 0 & 0 \cr \end{pmatrix} \; ,
\end{align}
and $U = u^2$ is defined in eq.~(\ref{eq:U}). The two couplings have been estimated as $f_v = F_V/M_V \simeq 0.20$ (from $\rho^0 \to e^+ e^-$) and $g_v = G_V/M_V \simeq 0.09$ (from $\rho \to \pi\pi$)~\cite{Bijnens:1992uz, DAmbrosio:2006xmn}. 

\subsection{\texorpdfstring{$\Delta S = 1$}{Delta S = 1}}
\label{sec:vectorDS1}
The $\Delta S=1$ hadronic effective Lagrangian in presence of external currents yields contributions to the Chiral Lagrangian which involve both pseudoscalar and vector mesons. The former, as discussed in ref.~\cite{Isidori:2005xm}, yield eq.~(\ref{LpsDS1}). The latter are more complicated because, at lowest order in the Chiral expansion, several independent operators are possible~\cite{Ecker:1992de}:
\begin{align}
{\cal L}_{\rm vect}^{\Delta S=1} &= \sum_i g_V^i K_i^V \; ,\\
K_1^V  &= \langle \Delta \{\hat V_{\mu\nu}, f_+^{\mu\nu} \} \rangle  \; ,\\
K_4^V  &= \epsilon_{\mu\nu\alpha\beta} \langle \Delta \{\hat V^{\mu\nu}, f_-^{\alpha\beta} \} \rangle  \; ,\\
K_5^V &= i \langle \Delta \{ \hat V_{\mu\nu} , [u^\mu,u^\nu] \} \rangle \; , \\
K_6^V &= i \langle \Delta  u_\mu \hat V_{\mu\nu}  u^\nu \rangle  \; ,\\
K_9^V &= \langle \partial_\nu \Delta \{u_\mu,\hat V^{\mu\nu}\} \rangle \; ,
\end{align}
where $\Delta = u \lambda u^\dagger$ and we included only operators even under Parity. 

The coefficients $g_V^i$ are very poorly known and rough estimates of their sizes can be extracted within various simplified models. For instance, the Weak Deformation model~\cite{Ecker:1992de} is based on the observation that  the Lagrangian ${\cal L}_{\rm ps}^{\Delta S=1}$ follows from the leading Chiral Lagrangian ${\cal L}_{\rm ps}$ via the substitutions:
\begin{align}
u_\mu &\to u_\mu + \{ u_\mu, \hat \Delta \} - \frac{2}{3} \langle u_\mu \hat \Delta \rangle \; ,\\
\Gamma_\mu &\to \Gamma_\mu + \frac{i}{2} \{ u_\mu, \hat \Delta \} - \frac{i}{3} \langle u_\mu \hat \Delta \rangle \; ,
\end{align}
where $\hat \Delta = G_8 F^2 \Delta$. Using the following relations:
\begin{align}
f_+^{\mu\nu} &= 2 i \Gamma^{\mu\nu} -\frac{i}{2} [u^\mu,u^\nu] \\
\Gamma^{\mu\nu} &= \partial^\mu \Gamma^\nu - \partial^\nu \Gamma^\mu + [\Gamma^\mu,\Gamma^\nu]
\end{align}
we can express $f_+^{\mu\nu}$ in terms of $u^\mu$ and $\Gamma^\mu$ and thus apply the Weak Deformation model transformation to ${\cal L}_{\rm vect}$ obtaining:
\begin{align}
\left[{\cal L}_{\rm vect}^{\Delta S=1}\right]_{\rm WD} &= 
\frac{G_8 F^2}{2\sqrt{2}} \left[ 2 f_V K_9^V + (g_V - \frac{f_V}{2}) (4 K_6^V - K_5^V ) \right]
\end{align}
implying that the only non-vanishing coefficients are
\begin{align}
[g_V^5]_{\rm WD} &= -\frac{1}{2\sqrt{2}}(g_V - \frac{f_V}{2} ) G_8 F^2 B_5 \; ,\\
[g_V^6]_{\rm WD} &= \frac{2}{\sqrt{2}}(g_V - \frac{f_V}{2} ) G_8 F^2 B_6\; ,\\
[g_V^9]_{\rm WD} &= \frac{1}{\sqrt{2}} f_V G_8 F^2 B_9\; , 
\end{align}
where we introduced bag parameters $B_{6,7,9}$ which parameterize the deviation of the matching coefficients $g_V^{5,6,9}$ from their values in the weak deformation model.

\begin{figure}[t]
\begin{center}
\includegraphics[height=0.2 \linewidth]{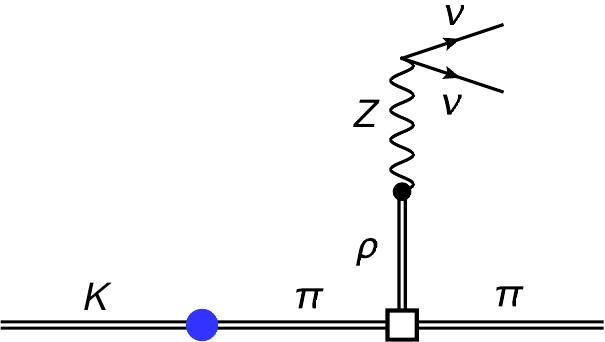}
\hfill
\includegraphics[height=0.2 \linewidth]{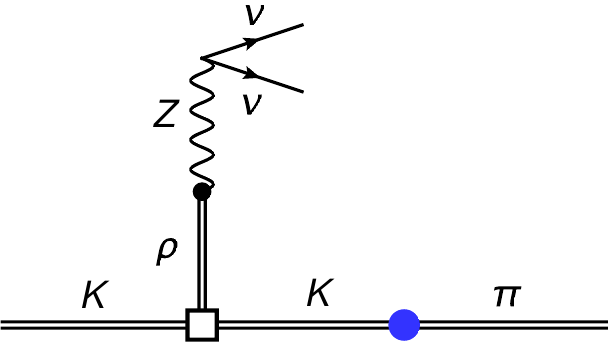}
\hfill
\includegraphics[height=0.2 \linewidth]{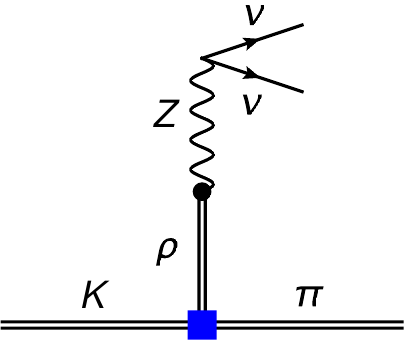}
\caption{Long distance contributions to $K^+\to \pi^+\nu\bar\nu$ in chiral perturbation theory. Empty squares, filled squares and filled circles represent insertions of ${\cal L}_{\rm vect}$, $\left[{\cal L}_{\rm vect}^{\Delta S=1}\right]_{\rm WD}$ and ${\cal L}_{\rm ps}^{\Delta S=1}$, respectively.}
\label{fig:longdistance-chipt}
\end{center}
\end{figure}
\begin{figure}[t]
\begin{center}
\begin{align}
\includegraphics[width=0.125 \linewidth]{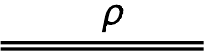}\;\;\;\;\;\;\;\;\;
&\;\; =  i\left[ \frac{g_{\mu\rho} g_{\nu\sigma}}{m_\rho^2} +\frac{g_{\mu\rho}q_\nu q_\sigma - g_{\mu\sigma} q_\nu q_\rho}{m_\rho^2 (m_\rho^2-q^2)} - (\mu\leftrightarrow \nu)\right]\\
\includegraphics[width=0.25 \linewidth]{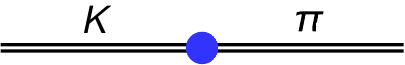}
&\;\; = -2 i F^2 G_8 \; p_K\cdot p_\pi\\
\includegraphics[width=0.25 \linewidth]{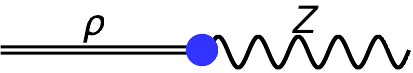}
&\;\; = F_V g_2 \frac{\cos 2\theta_W}{4\cos\theta_W} ( g_{\alpha\nu} p_\mu - g_{\alpha\mu} p_nu) \\
\includegraphics[width=0.25 \linewidth]{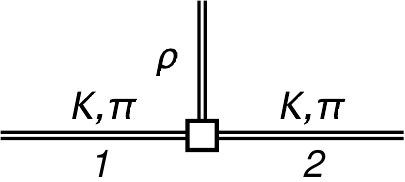}
&\;\; =i  \frac{G_V}{F^2 \alpha_{K,\pi} } (p_1^\mu p_2^\nu-p_1^\nu p_2^\mu) \quad\quad [\alpha_K=2, \alpha_\pi =  1 ]\\
\includegraphics[width=0.25 \linewidth]{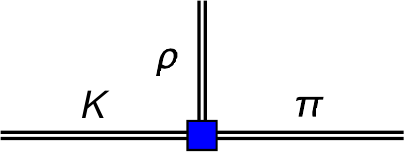}
&\;\; = i \frac{2\sqrt{2} (2 g_5^V-g_6^V)-g_9^V}{4 F^2} (p_K^\mu p_\pi^\nu - p_K^\nu p_\pi^\mu )
\end{align}
\caption{Feynman rules originating from the Lagrangians ${\cal L}_{\rm vect}$, $\left[{\cal L}_{\rm vect}^{\Delta S=1}\right]_{\rm WD}$ and ${\cal L}_{\rm ps}^{\Delta S=1}$.}
\label{fig:feynman}
\end{center}
\end{figure}

\section{Calculation of long distance contributions}
\label{sec:calculation}
Contributions to the $K^+\to \pi^+\nu \bar\nu$ mediated by vector mesons are shown in figure~\ref{fig:longdistance-chipt}. The relevant Feynman rules originate from the Lagrangians ${\cal L}_{\rm ps}^{\Delta S=1}$, ${\cal L}_{\rm vect}$ and $\left[{\cal L}_{\rm vect}^{\Delta S=1}\right]_{\rm WD}$ and are summarized in figure~\ref{fig:feynman} (see, for instance, ref.~\cite{Ecker:1988te} for a discussion of Feynman rules for anti-symmetric fields). We adopt the standard Breit-Wigner parameterization of the $\rho$ meson propagator.

The $K^+\to \pi^+\nu \bar\nu$ amplitude reads:
\begin{align}
{\cal A} = \;& \frac{G_F}{\sqrt{2}} \frac{\alpha \lambda^5}{2\pi \sin^2\theta_W} 
(p_K^\mu + p_\pi^\mu) \sum_{\ell=e,\mu,\tau} \bar u_{\nu_\ell} \gamma_\mu (1-\gamma^5) v_{\nu_\ell} \; [\delta P_{c,u}]_\rho (q^2) \; , \\
[\delta P_{c,u}]_\rho (q^2) = \; & 
\frac{2 \pi^2 \cos 2\theta_W }{\lambda^4} g_8 \frac{F_V G_V}{m_\rho^2}  \frac{m_\rho^2}{m_W^2}  
\frac{q^2}{q^2-m_\rho^2+i m_\rho \Gamma_\rho} \Bigg[ \frac{m_K^2-2 m_\pi^2}{m_K^2-m_\pi^2}  \nonumber\\
& -\left[ \left(1-\frac{F_V}{2G_V} \right) (B_5 + 2 B_6)  + \frac{1}{2\sqrt{2}} \frac{F_V}{G_V} B_9 \right] \Bigg] \;,   \label{eq:deltaPrho}\\
[\delta P_{c,u}]_\rho^{\rm WD} (q^2)\simeq \; &
\frac{\pi^2 \cos 2\theta_W }{\lambda^4} g_8 \frac{F_V G_V}{m_\rho^2}  \frac{q^2}{m_W^2}
\end{align}
where in the last line we set $B_i \to 1$, rounded the numerical value of the square bracket to $ m_1/2$.
\begin{figure}[t]
\begin{center}
\includegraphics[width=0.99 \linewidth]{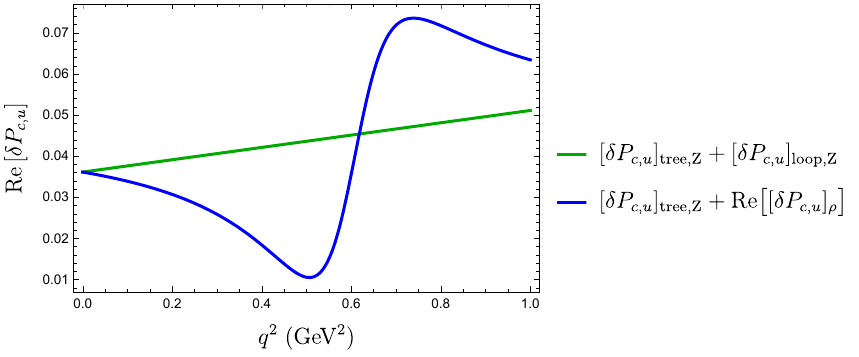}
\includegraphics[width=0.99 \linewidth]{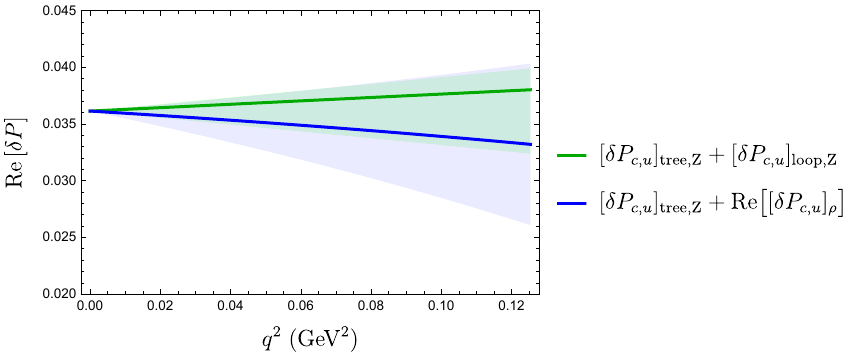}
\caption{Tree level, one loop and vector meson contributions to the matrix elements $\delta P_{uc}$ of the operators $O_{1,2}^u$ within ChiPT. The vector meson contribution has a small imaginary part. The left panel shows a wide phase space region which includes the $\rho$ resonance; the right panel is limited to the the $K^+\to \pi^+ \nu\bar\nu$ phase space.}
\label{fig:dP}
\end{center}
\end{figure}

In figure~\ref{fig:dP} we show the comparison between $Z$ mediated contributions to the real part of $\delta P_{c,u}$ calculated using pseudoscalar meson loops (without the inclusion of unknown counterterms~\cite{Isidori:2005xm}) and in terms of $\rho$ meson exchange. The former are taken from eqs.~(60) and (61) of ref.~\cite{Isidori:2005xm} and the latter from eq.~(\ref{eq:deltaPrho}) with all $B$ parameters set to 1. In both cases we add the tree level contributions, given in eq.~(\ref{ps-tree}), stemming from the complete leading power matching of the weak effective Hamiltonian onto the Chiral Lagrangian. In the left panel we show the differential amplitude for an extended range of $q^2$ to display the typical behavior encountered around the $\rho$ resonance. In the right panel we restrict the range to the $K\to \pi \nu\bar \nu$ phase space, $q^2 < (m_K-m_\pi)^2$, and include a rough estimate of theoretical uncertainties. The latter have been obtained by allowing a range of $\pm 2 [\delta P_{c,u}]_{{\rm loop},Z}$ around the leading contribution $[\delta P_{c,u}]_{{\rm tree},Z}$ and by varying all $B$ parameters in the $[0,2]$ range. We note that, while the pseudoscalar loop and vector meson exchange calculations yield compatible results, the latter are a complete calculation (albeit up to the weak deformation model assumption) of these effects. 

After phase space averaging we obtain the following estimate for the real part of the $\rho$-exchange contribution (the imaginary part is much smaller):
\begin{align}
\langle \text{Re} \big[ [\delta P_{c,u}]_{\rho} \big]\rangle = \left(-1.0 \pm 2.5\right) \times 10^{-3} \; .
\end{align}

In order to provide a complete estimate of light-quark loops to the amplitude we need to add contributions stemming from charged currents which have been calculated (again up to unknown counterterms) in ref.~\cite{Isidori:2005xm}. Note that, in this case, it is not possible to use resonance Chiral Perturbation Theory to obtain an estimate of the missing counterterms. After including these $W$ exchange contributions with a 100\% uncertainty we obtain the following estimate:
\begin{align}
\langle \text{Re} [\delta P_{c,u}] \rangle = \left(3.1 \pm 0.3_{\rho} \pm 0.3_{W}\right) \times 10^{-2} \; .
\label{eq:finalresult}
\end{align}%
Note that the central value is essentially determined by the complete leading result in eq.~(\ref{eq:GinoUpdated}) for which we find $2.9\times 10^{-2}$. 

\section{Kaon Unitarity Triangle: current status and projections}
\label{sec:kut}
In ref.~\cite{Lehner:2015jga} we considered a unitarity triangle fit based almost exclusively on kaon observables. The ingredients of the fits are $\varepsilon_K$, $\varepsilon^\prime_K/\varepsilon_K$, ${\rm BR} (K^+\to \pi^+\nu\bar\nu)$, ${\rm BR} (K_L \to \pi^0\nu\bar\nu)$ and $V_{cb}^{}$ from inclusive and exclusive $b\to c \ell\bar\nu$ decays. We refer to ref.~\cite{Lehner:2015jga} for all relevant formulae and the definition of the various quantities. 

The rationale behind this strategy is based on the observation that the standard unitarity triangle (SUT) fit is currently dominated by measurements of the three angles in various $B$ decays and of the $B_q-\bar B_q$ (q=d,s) mass differences and presents a picture
in good agreement with the CKM mechanism (see, for instance, the most recent analyses of the UTfit~\cite{UTfit:2022hsi} and CKMfitter~\cite{Qian:2023eok} collaborations). Assuming that New Physics enters the picture via contributions to the Kaon sector, the best way to isolate it and to study the impact of future experimental and theoretical progress is to compare the Standard (SUT) and Kaon (KUT) Unitarity Triangles. In particular, $\varepsilon^\prime_K $ and $K_L\to \pi^0\nu\bar\nu$ provide constraints on the CP violating phase that are parameterized by $\bar \eta$ and that are expected to improve in the near future. The $K^+\to \pi^+ \nu\bar\nu$ branching ratio has been already measured with reasonable accuracy, the resulting constraint in the $(\bar\rho, \bar \eta)$ plane is presently dominated by experimental uncertainties and reducing the residual theoretical error on the quantity $\delta P_{c,u}$ is one of the purposes of this paper. Finally, theoretical and experimental progress on the indirect CP violating quantity $\varepsilon_K$ and the extraction of $|V_{cb}|$ from semileptonic decays is not expected to yield a sizable impact on Unitarity Triangle analyses. 

\begin{figure}[t]
\begin{center}
\includegraphics[width=0.8 \linewidth]{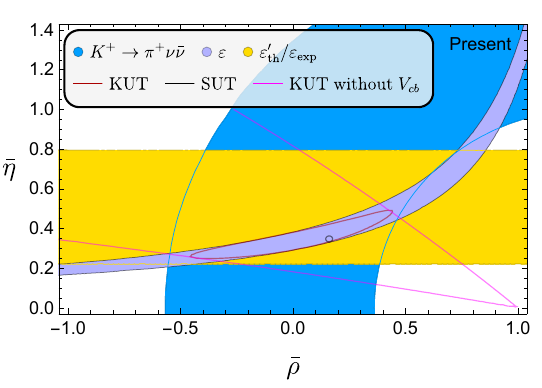}
\includegraphics[width=0.8 \linewidth]{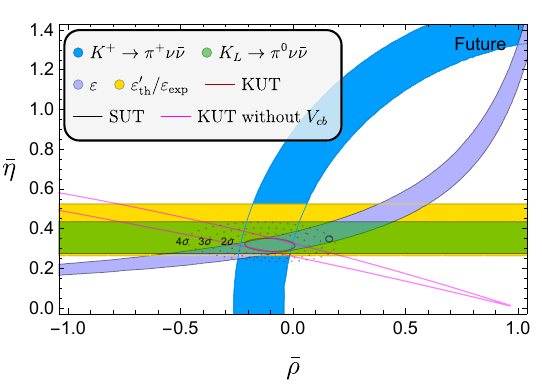}
\caption{
Top panel: current status of the Standard (SUT) and Kaon (KUT) Unitarity Triangles. Bottom panel: impact of improved calculations of ${\rm Im} A_{0,2}$ from lattice QCD and of expected measurements of charged (NA62) and neutral (KOTO) $K\to \pi \nu\bar\nu$ branching ratios on the Kaon Unitarity Triangle. The three dotted contours are the $2\sigma$--$4\sigma$ KUT contours, respectively.
}
\label{fig:kfitNoVcb}
\end{center}
\end{figure}

Since the analysis of ref.~\cite{Lehner:2015jga} the RBC-UKQCD collaboration substantially improved the calculation of ${\rm Im} A_0$~\cite{RBC:2020kdj}, the NA62 collaboration presented an updated measurement of the $K^+ \to \pi^+ \nu\bar\nu$ branching ratio using data from 2016-2018~\cite{NA62:2021zjw} and the KOTO collaboration presented results based on data collected in 2015~\cite{KOTO:2018dsc}, 2016-2108~\cite{KOTO:2020prk} and 2019-2021~\cite{KOTOMoriond2024}:
\begin{align}
{\rm Im} A_0 &= -6.98(0.62)(1.44) \times 10^{-11} \; {\rm GeV} \; , \label{ImA0} \\
{\rm BR} (K^+\to \pi^+\nu\bar\nu)_{\rm exp} &= ( \left. 10.6^{+4.0}_{-3.4}\right|_{\rm stat} \pm 0.9_{\rm syst} )\times 10^{-11} \; , \label{k2pi+nn} \\
{\rm BR} (K_L\to \pi^0\nu\bar\nu)_{\rm exp} & < 2.0 \times 10^{-9} \;\; @ 90\%{\rm C.L.} \; .
\end{align} 
We present the current status of the Kaon Unitarity Triangle fit in the upper panel of figure~\ref{fig:kfitNoVcb}. The contours corresponding to $\varepsilon_K$, $\varepsilon^\prime_K/\varepsilon_K$ and ${\rm BR}(K^+\to \pi^+\nu\bar\nu)$ (currently the upper limit on ${\rm BR}(K_L\to \pi^0\nu\bar\nu)$ provides too loose constraints) are obtained with the inclusion of experimental information on $|V_{cb}^{}|$ from inclusive and exclusive semileptonic $b\to c \ell\nu$ decays; in fact, without information on parameter $A$ in the Wolfenstein parameterization of the CKM matrix it is not possible to place constraints in the $(\bar\rho,\bar\eta)$ plane stemming from any one of these observables. The combined fit is compatible with the Standard Unitarity Triangle fit which is obtained by including all relevant $B$ physics observables. Note that we calculate $\varepsilon_k$ using the framework presented in ref.~\cite{Brod:2019rzc} in which NNL and NNLL QCD corrections are parameterized in terms of the two quantities $\eta_{tt}$ and $\eta_{ut}$ whose uncertainties are largely uncorrelated. The actual results for the Wolfenstein parameters are: $\bar\rho = -0.03 \pm 0.29$ and $\bar\eta = 0.340 \pm 0.071$. 

We also present (magenta contour) the region allowed by the combination of $\varepsilon_K$, $\varepsilon^\prime_K/\varepsilon_K$ and ${\rm BR}(K^+\to \pi^+\nu\bar\nu)$ {\em without} making use of information on $|V_{cb}^{}|$. This contour is in the shape of a long flat ellipsis which extends to the point $(\bar\rho,\bar\eta)=(1,0)$. The puzzling fact that $\bar\eta =0$ is allowed while fitting two observables that vanish exactly in that limit ($\varepsilon_K$ and $\varepsilon^\prime_K$) can be understood by observing that the ratio $(\varepsilon_{\rm th}^\prime/\varepsilon_{\rm th})$ depends only on the Wolfenstein parameters $\bar \rho$ and $A$ and is independent of $\bar \eta$. For $\bar\rho=1$ this ratio and the $K^+\to \pi^+ \nu\bar \nu$ branching ratio are also independent of $A$ implying that the $(\bar\rho,\bar\eta)=(1,0)$ point is always allowed.

In the lower panel of figure~\ref{fig:kfitNoVcb} we present a possible future scenario~\cite{Anzivino:2023bhp, DAmbrosio:2023irq} in which we assume that the $K_L\to \pi^0\nu\bar\nu$ branching ratio is measured at the 27\% level (as projected by the KOTO~II experiment~\cite{Nanjo:2023xvj}), the $K^+\to\pi^+\nu\bar \nu$ branching ratio is determined with a 5\% uncertainty (after the completion of the experimental programs at NA62)                                                                                                                                                                                                     
and the matrix element ${\rm Im} A_0$ is calculated at the 10\% level by the RBC collaboration. We note that the current calculation of the quantity ${\rm Im} A_0$~\cite{RBC:2020kdj} is performed using $G$-parity boundary conditions(GPBC)  on the lattice and that the upcoming analysis is going to be based on periodic  boundary conditions (PBC) ~\cite{RBC:2023ynh} which seem much less computationally intensive.  


In order to exemplify the ability of this analysis to disentangle new physics in the $B$ and $K$ sector we assume that the central values for the future measurement of the $K^+\to\pi^+\nu\bar \nu$ branching ratio and of the calculated matrix element ${\rm Im} A_0$ will remain at their current values. The results for the Wolfenstein parameters in this future scenario are $\bar\rho = -0.100 \pm 0.074$ and $\bar\eta = 0.316 \pm 0.022$. We see that in this scenario it is conceivable to establish a tension between the Kaon and Standard Unitarity Triangle fits between the 3 and 4$\sigma$ level (dotted contours). If no use of charmless semileptonic decays is made, the future sensitivity reduces to the 2-3$\sigma$ level.

%
%
%
%

%
%

%

\section{Conclusions}
\label{sec:conclusions}
In this paper we pursue a simple approach to the calculation of long-distance contributions to the rare decay $K^\pm \to \pi^+ \nu\bar\nu$. These terms are responsible for the largest non-parametric contribution to the theoretical prediction and stem from matrix elements which, at the quark level, involve up-quark loops. A proper calculation of these effects can only be performed within lattice QCD; unfortunately the technical challenges involved are considerable and a calculation with reasonably small uncertainties is not expected for several years. 

Currently, the only avenue to an estimate of these corrections is to match the relevant $\Delta S=1$ operators onto the Chiral Lagrangian. The main difficulty with this approach is the poor knowledge of the relevant matching conditions. While the leading power matching involving pseudoscalar mesons are known, the same is not true for the subleading power counterterms required to perform the calculation at one loop (in ChiPT). 

In this paper we propose to use vector meson dominance to obtain an independent estimate of the above mentioned loop effects. The starting point is the inclusion of vector mesons into the Chiral Lagrangian within the framework of Resonant Chiral Perturbation Theory. The advantage of this approach is that all required $\Delta S=1$ matching conditions can be estimated using various approaches, like the weak deformation model or factorization. The results we obtain, presented in eq.~(\ref{eq:finalresult}) and in figure~\ref{fig:dP}, are compatible with the state-of-art result given in eq.~(\ref{eq:GinoUpdated}):
\begin{align}
\langle \delta P_{c,u} \rangle &= 
\begin{cases}
(3.1 \pm 0.3_{\rho} \pm 0.3_{W} ) \times 10^{-2}  & \text{vector meson exchange,} \\
(3.0 \pm 1.5 ) \times 10^{-2}  & \text{pseudoscalar meson loops~\cite{Isidori:2005xm}.} \\
\end{cases}
\end{align}

We conclude with an update on the current status and future prospects of the Kaon Unitarity Triangle, namely a CKM fit which involves (almost) exclusively observables in the Kaon sector: $\varepsilon_K$, $\varepsilon_K^\prime/\varepsilon_K$, ${\rm B} (K^\pm \to \pi^\pm \nu\bar \nu$ and $K_L \to \pi^0 \nu\bar\nu$.  The main motivation for studying this fit is that new physics scenarios which result in sizable contributions mostly confined within the Kaon sector (which are known to happen in several beyond-the-SM models) might not be visible in Standard Unitarity Triangle fits given the extreme accuracy of many observables entering the latter. 

While traditionally the importance of the gold-plated mode, $K_L \to \pi^0 \nu \bar \nu$ for the KUT is emphasized, given that this very challenging measurement still has perhaps a decade to get there, we reassert the importance of $\varepsilon_K^\prime/\varepsilon_K$ via the lattice as it appears that the uncertainties can be reduced to the level  of 10\% in the next few years. Efforts in constraining $K_L \to \pi^0 \nu \bar \nu$  via $ K^0 \to \pi^0 l^+ l^-$ are also emphasized.

\section*{Acknowledgements}
We thank Robert Szafron and Gilberto Colangelo for discussions in the early (and late) stages of this work.

\bibliography{references}{}
\bibliographystyle{JHEP} 

\end{document}